\documentclass[12pt,fleqn,usenatbib]{mnras}
\usepackage{newtxtext,newtxmath}
\usepackage[T1]{fontenc}
\usepackage{ae,aecompl}
\usepackage{graphicx}
\usepackage{amsmath}
\usepackage{amssymb}
\usepackage{enumitem}
\usepackage{multirow}
\usepackage{gensymb}
\usepackage{threeparttable}
\usepackage{times}

\graphicspath{{images/}}

\title[LIM with Neural Networks]{Deconfusing intensity maps with neural networks}
\author[D. N. Pfeffer et al.]{
Daniel N. Pfeffer,$^{1}$\thanks{Email: dpfeffe2@jhu.edu (DNP)}
Patrick C. Breysse,$^{2}$
and George Stein,$^{2,3}$
\\
$^{1}$Department of Physics and Astronomy, Johns Hopkins University, Baltimore, MD 21218, USA\\
$^{2}$Canadian Institute for Theoretical Astrophysics, University of Toronto, 60 St. George Street, Toronto, ON, M5S 3H8, Canada \\
$^{3}$Department of Astronomy \& Astrophysics, University of Toronto, 50 St. George St., Toronto, ON, M5S 3H4, Canada
}

\date{Accepted XXX. Received YYY; in original form ZZZ}

\pubyear{2019}

\begin{document}
\label{firstpage}
\pagerange{\pageref{firstpage}--\pageref{lastpage}}
\maketitle

\begin{abstract}
Line intensity maps (LIMs) are in principle sensitive to a large amount of information about faint, distant galaxies which are invisible to conventional surveys. However, actually extracting that information from a confused, foreground-contaminated map can be challenging.  In this work we present the first application of convolutional neural network (CNN) to directly determine the underlying luminosity function of a LIM, including a treatment of extragalactic foregrounds and instrumental noise. We apply the CNN to simulations of mock Carbon Monoxide (CO) line intensity maps similar to those which will be produced by the currently-active COMAP experiment. We evaluate the trained CNN on a number of noise scenarios in order to determine how robust the network predictions are for application to realistic data. We find that, in the ideal case where the mock data capture all of the features of the real data, the CNN performs comparably to or better than conventional analyses.  However, the network's accuracy degrades considerably when tested on signals and systematics outside of those it was trained on.  For both intensity mapping and cosmology as a whole, this motivates a broad-based study of whether simulated data can ever be generated with sufficient detail to realize the enormous potential of machine learning methods.
\end{abstract}


\section{Introduction} \label{sec:int}
%\gs{we should use citet{} and citep{} throughout for MNRAS format. e.g. from their latex template: cite relevant earlier studies in the field by \citet{Others2013},
%and describes the problem the authors aim to solve \citep[e.g.][]{Author2012}.}

A significant experimental effort is underway to study the high-redshift universe with line intensity mapping (LIM).  LIM experiments seek to probe galaxy evolution and large-scale structure without resolving individual emitting sources.  Instead, these surveys seek to map the aggregate emission of a single spectral line over cosmological scales \citep[see][for a review]{Kovetz2017}.  Because the target emission comes from narrow spectral features, one can observe at may closely-spaced frequency bands to map the distribution of emitters in three dimensions.  Intensity maps can therefore access a large number of spatial modes for large-scale structure measurement, and can study the statistical properties of large numbers of galaxies which are too faint to detect individually.  

This great potential science output has spurred the creation of LIM surveys targeting a number of different spectral lines.  The first line targeted was the 21 cm spin-flip transition in neutral hydrogen, which has been long known as a powerful probe of large-scale structure and reionization \citep[][and references therein]{Pritchard2012}.  The 21 cm intensity mapping signal has been detected in cross-correlation by a pair of surveys \citep{Masui2013,Anderson2018}, and a number of other surveys have been completed or are in progress across a wide swath of cosmic history \citep{Tingay2013,vanHaarlem2013,Bandura2014,Ali2015,Xu2015,Newburgh2016,DeBoer2017}.  Recently, though, there has been a surge of interest in using other lines for intensity mapping.  Different lines trace different processes and different phases in the interstellar- and intergalactic media, and also have different experimental systematics.  There are tentative detections of intensity maps of Lyman-$\alpha$ \citep{Croft2018}, \ion{C}{ii} \citep{Pullen2018,Yang2019}, and CO(1-0) \citep{Keating2016}, and experiments are underway or proposed to make definitive measurements of these and other lines \citep{Kovetz2019,Hill2008,Crites2014,Dore2014,Bower2016,Cooray2016,Li2016,Aguirre2018,Lagache2018,Padmanabhan:2017ate,Stacey2018}.

With this degree of experimental investment, it is important to develop the necessary theory and analysis tools to interpret the results of these surveys.  Several challenges remain with this task.  Line intensities are determined by complex and highly nonlinear gas physics which can only be captured by quite sophisticated models \citep[see, e.g.][]{Popping2019}.  As a result of this complexity, many works focus on constraining intermediate statistical properties of the target galaxies, typically either a relationship between halo mass and line luminosity \citep[][]{Li2016} or a line luminosity function \citep[][]{Breysse2016}.  

This goal is made more difficult by the unresolved nature of intensity mapping data, as we must construct statistics which link the intensity field and the underlying galaxy distribution.  For example, the power spectrum of a map can be used to determine the first two moments of the target luminosity function \citep{Lidz2011,Breysse2019}.  Further detail can be obtained using, for example, the one-point statistics of a map \citep{Breysse2016,Ihle2019}.  However, these statistics may not suffice to extract all of the useful information from a confused, highly non-Gaussian intensity map, especially given that the target line is rarely the only or the most dominant source of emission.  These statistics must be modified and lose additional information due to the near-guaranteed presence of foreground contamination, both from local Milky Way emission \citep[see, e.g.][]{Morales2010,Wolz2015} and from extragalactic sources \citep{Sun2018,Switzer2019}.  In light of the difficulty of measuring a luminosity function from a contaminated map, it may be useful to consider different analysis approaches.  In this work, we will explore possibilities for applying machine learning methods to intensity maps.

In recent years, machine learning (ML) methods have shown to be very useful for a variety of applications in the field of cosmology\footnote{\label{note1}comprehensive list at https://github.com/georgestein/ml-in-cosmology}, and will continue to contribute significant cosmological insights over the following decade and beyond \citep{mldecadal}. The utility of machine learning methods emerges from their ability to find patterns in data, and, in many cases, to relate these patterns to higher-level information about samples from the data set. For example, when applying a neural network to solve the classic computer vision application of classifying handwritten digits, the network learns patterns in the spatial distribution of the two dimensional pixel intensities in order to predict the higher-level class (an integer between 0-9) of an individual digit sample drawn from the test set. Similarly, in cosmology, machine learning can be used on data from simulations, observations, or (possibly) a combination of the two to predict cosmological or astrophysical parameters \citep{ml:param1, ml:param2, ml:param3, ml:param4, ml:param6, ml:param5}, perform model discrimination \citep{ml:model1, ml:model2}, augment simulations and create synthetic data \citep{ml:sim1, ml:sim2, ml:sim3, ml:sim4, ml:sim5, ml:sim6, ml:sim7}, identify structures and predict their properties \citep{ml:structureclassification, ml:massmeasurement, ml:galaxyshapes}, or to reconstruct initial conditions \citep{ml:reconstruct1}, among many other applications.

Convolutional neural networks (CNNs) are a common class of deep learning first proposed in \citet{cnn1, cnn2} and popularized by the state-of-the-art classification results of \citet{alexnet}. CNNs are best designed to process data that come in the form of multiple arrays; the most common example being the three colour channels, or RGB pixel arrays, of two dimensional images, and they have had success in a wide variety of detection, segmentation, and recognition applications. A CNN transforms from an input N-dimensional array to e.g. a prediction of which class the array belongs to. Convolutional layers consist of a number of filters, each containing of a set of trainable weights (determined through backpropagation \citep{backprop}) which are applied to a series of local patches of the previous layer. This allows the network to detect local features in the previous layer, and the network can learn higher-level information in each succeeding level.

CNNs are therefore particularly suited to problems in cosmology that require environmental multi-scale information to solve. For example, it is well known that the large-scale structure of the universe is defined by the network of clusters (small), filaments (elongated), and voids (large), of the cosmic web \citep{cosmicweb}, each with differing physical scales. One may then hope that if, for example, the observable signal from clusters is related to the surrounding environment: the first levels of a CNN will extract features relevant to the scale of a cluster, following levels will focus on features relevant to broader cluster environments, succeeding levels will add features related to the large-scale distribution of matter in the universe, and this multi-scale information will be combined in the final levels to make a prediction (see \citet{2013arXiv1311.2901Z} for investigations into which input stimuli excite individual feature maps at any
layer in a model).  

Although powerful in theory, many observational cosmological applications of supervised machine learning still have obstacles to overcome before becoming competitive with alternative methods. Astronomy is a field of observation, and contains little to no possibility of experimentation. Additionally, the exact amplitude, extent, and spectral evolution of many of the cosmic signals that we are attempting to detect in fields such as intensity mapping are presently unknown, and labeled observational data sets are in many cases theoretically difficult or impossible to acquire. The field has therefore been focused more on  studies performed on synthetic data to determine the general viability of machine learning methods to extract cosmic signals. 

The use of synthetic data is not uncommon in machine learning applications \citep{syntheticdata1, syntheticdata2}, but is generally used to augment small existing data sets and is followed by additional training on the true labeled data. Currently this is not possible for many observational cosmology applications, so we must hope that: a) the synthetic data perfectly reproduces reality, and that by training on synthetic observations and using the network to predict on a true observation of our universe therefore produces no biases or uncertainties (unlikely), b) the network is sufficiently robust to any differences between the synthetic and real data (e.g. unaccounted for instrument errors, unknown foreground contamination, etc.), and any biases and uncertainties are well understood, or c) labelled training data from cosmological measurements becomes plentiful enough to rely on, and machine learned methods outperform traditional ones. In this work we rely purely on synthetic data, and focus on scenario b). By first studying the ideal case of perfectly known cosmic signal and instrument noise, and then extending to add unknown foregrounds and noise to mock up a real observation, we can shed light on the true ability of a network trained on synthetic data to measure cosmic signals.

This work is not the first attempt train CNNs on simulated intensity maps. Previous works primarily study maps of the 21 cm spin-flip transition in neutral hydrogen \citep{Gillet2019,Hassan2018,Hassan2019,Zamudio2019}.  Many of these works focus on the Epoch of Reionization (EoR), where the signal is dominated by emission from the intergalactic medium which is gradually becoming ionized by emission from the first galaxies.  In this work. we consider a different regime, where line emission primarily comes from within individual galaxies.  In this case, individual sources are typically small compared to instrument resolution, so there is a well-defined line luminosity function that we can seek to constrain.  We focus on CO intensity mapping as opposed to HI, which gives insight into the molecular phase of the high-redshift ISM.  We seek to be as model-agnostic as possible in our predictions by forecasting constraints on the value of the luminosity function in different bins rather than constraining a specific parameterized model.

For our fiducial survey, we consider a map of the CO(1-0) line at redshift $z\sim3$ made by the CO Mapping Array Pathfinder (COMAP) experiment \citep{Li2016}, currently taking data at the Owens Valley Radio Observatory.  The CO luminosity function probes the abundance of molecular gas in high-redshift galaxies.  As stars form from molecular gas, the CO luminosity is an important probe of the broader galactic ecosystem \citep[see reviews by, e.g.][]{Bolatto2013,Carilli2013,Heyer2015}.  As stated previously though, our ML methods will be directly relevant to any line which is emitted by a population of discrete sources.

We demonstrate that for our fiducial model the neural network we create can recover the luminosity function from a  CO intensity map with accuracy comparable to that of conventional methods.  The accuracy only degrades slightly when contamination is introduced from instrument noise and uncleaned foregrounds.  However, when testing on models on the fringes of our training space, or on models which were not trained at all, we find that the network sometimes outputs substantially inaccurate results.  These findings demonstrate that, while machine learning methods have great potential for this type of data analysis, care must be taken when using synthetic data to analyze real observations.

This paper is organized as follows.  In Section \ref{sec:maps}, we describe how we generate our data and training set.  In Section \ref{sec:nn}, we describe the CNN that we will train.  In Section \ref{sec:res}, we test the accuracy of our CNN on different scenarios for underlying LIM, noise and foregrounds.  In Section \ref{sec:disc}, we discuss the strengths and weaknesses of our CNN.  We conclude in Section \ref{sec:conc}.  Throughout this work we assume a cosmology consistent with \citet{Planck2018} with $\Omega_m$ = 0.286, $\Omega_\Lambda$ = 0.714, $\Omega_b$ = 0.047, h = 0.7, $\sigma_8$ = 0.82, and $n_s$ = 0.96.

\section{Simulated Maps} \label{sec:maps}
Very little actual CO intensity mapping data currently exists, so as stated above we have to resort to synthetic data to train our neural network.  For this purpose, we use a set of simulated CO line intensity observations constructed by coupling the $\rm{L_{CO}(M_{halo})}$ model of \citet{Li2016} to dark matter halo catalogues created using the Peak Patch method \citep{peakpatch}.  Our goal is to train a neural network that can take one of these simulated maps as input and output a list of galaxy abundances at given CO luminosities. In addition to the intrinsic CO signal we add various noise sources to our maps, including the thermal white noise expected from the COMAP experiment, possible point source foregrounds with continuum spectra, and `geometric' noise from crude approximations of typical instrumental scan strategies.  Our simulations are not intended to fully reproduce the range of possible signals and noise in a real CO experiment.  Rather they are meant to explore how a CNN-based analysis might perform in a variety of conditions.

Table \ref{tab:comap} lists the experimental parameters we use for our mock LIMs.  We represent our generated LIMs as three dimensional arrays of size 64x64x100. Each element records the total intensity measured at that location in the map. The first two dimensions are spatial dimensions on the sky representing a 1.5\textdegree$\times$1.5\textdegree survey field, while the third dimension carries the spectral information.  Although COMAP is designed with 512 frequency channels, we intentionally degrade the frequency resolution of our mocks down to 100 channels due to memory considerations.  The 64x64 maps oversample the COMAP beam somewhat, so we can add in the effects of COMAP beam smoothing by convolving each slice of our maps with a $4'$ Gaussian filter. For a given spectral line, the observed frequency directly determines the emission redshift, so the third dimension in our maps represents the redshift (or distance, given a cosmological model) of CO emitters along the line of sight. In all of the following, a ``voxel" refers to a single element in a three-dimensional map, and a ``pixel" refers to all of the voxels along a line of sight when the first two dimensions (position on the sky) are kept constant.

\begin{table}
    \centering
    \caption{Experiment setup for COMAP Phase one.}
    \begin{tabular}{c|c}
      \hline		
      Parameter & Value \\
      \hline  
      Beam FWHM (armin) & 4 \\
      Frequency Band (GHz) & 26-34 \\
      Redshift coverage  & 2.4-3.4 \\
      Channel width (MHz) & 15.6 \\
      Noise per 16 armin\(^2\) voxel (\(\mu\)K) & 11 \\
      Field size (deg\(^2\)) & 2.25 \\
      \hline
    \end{tabular}
    \label{tab:comap}
\end{table}

\subsection{Dark Matter Simulations} \label{sec:sims}
We generated the large ensemble of dark matter halo catalogues required to train our CNN using the Peak Patch method, a fully predictive initial-space algorithm to quickly generate dark matter halo catalogues in large cosmological volumes \citep{peakpatch}. 

To cover the full redshift range of the COMAP experiment ($z=2.4-3.4$), with no repetition of structure, the simulation box size was (1140 Mpc)$^3$ (comoving) and used a cubic lattice of 4096$^3$ particles. This achieves a minimum halo mass of 2.5$\times$10$^{10}$M$_\odot$ [M$_{200,M}$], comparable to values typically assumed for the minimum mass of a CO-emitting halo \citep{Lidz2011,Li2016}, and when projected onto the sky results in a 9.6\textdegree$\times$9.6\textdegree\ field.  We then separate the 9.6\textdegree$\times$9.6\textdegree\ area into multiple 1.5\textdegree$\times$1.5\textdegree\ patches to match the size of a COMAP field.  Each 1.5\textdegree$\times$1.5\textdegree\ patch we use does not overlap with any other to minimize nonphysical correlations in our training data.

The efficiency of the Peak Patch method allowed for 161 independent full-size realizations in 82,000 CPU hours. The resulting halo catalogues contain roughly 54 million halos, each with a position, a velocity, and a mass. Peak Patch has the ability to simulate continuous light-cones on-the-fly, so stitching snapshots together was not required to create the light-cone. The dark matter halo catalogues were additionally mass corrected by abundance matching along the light-cone to \citet{tinker2008}.

\subsection{CO Modelling}
\label{sec:CO}
The peak-patch simulations described above give us a map of dark matter halos, which we can turn into a CO intensity map by assuming a CO luminosity-halo mass relation.  For this purpose, we adopt the model of \citet{Li2016} which we briefly summarize here.

The model is defined by empirical parametric relations between the halo mass $\rm{M_{halo}}$, star formation rate (SFR), infrared (IR) luminosity $\rm{L_{IR}}$, and the CO luminosity $\rm{L_{CO}}$, in the following chain:
\begin{equation*}
    \rm{M_{halo} \xrightarrow{A} SFR \xrightarrow{B} L_{IR} \xrightarrow{C} L_{CO}}.
\end{equation*}

\begin{enumerate}[label=\Alph*:]
    \item The star formation rate of a given dark matter halo is obtained by using the results of  \citet{behrooziB, behrooziA}, which
empirically quantified the average stellar mass history of dark matter halos as a function of halo mass and redshift. A log-normal scatter of $\rm{\sigma_{SFR}}$ is added to describe the scatter about the mean value.
\item IR luminosities are given through the relation
    \begin{equation}
        SFR = \delta_{MF} \times 10^{-10} L_{IR},
    \end{equation}
where the SFR is in units of $\rm{M_\odot\ yr^{-1}}$ and $\rm{L_{IR}}$ is in units of $L_\odot$.
\item CO luminosities are obtained from the total infrared luminosity, assuming a power-law relation of the form 
\begin{equation}
    \rm{log}\ L_{IR} = \alpha \rm{log}\ L_{CO}' + \beta,
\end{equation}
where $\rm{L_{CO}'}$ is in units of $\rm{K\ km\ s^{-1}}$, which we then convert in our final maps to brightness temperature in $\rm{\mu K}$. A second log-normal scatter of $\rm{\sigma_{CO}}$ is also added to describe the scatter about this mean value.
\end{enumerate}

This model therefore contains five free parameters: three parameterizing the mean relations \{$\delta_{MF}, \alpha, \beta$\} and two parameters describing the scatter about the mean, \{$\sigma_{SFR}$, $\sigma_{L_{CO}}$\}. In this work we fix the scatter about the mean to the fiducial values of \citet{Li2016}, \{$\sigma_{SFR}$, $\sigma_{L_{CO}}$\} = \{0.1, 0.1\}, while varying the others in the training step.  We do not train our CNN to predict these model parameters as one may do with a Markov Chain Monte Carlo (MCMC) or similiar analysis.  Our CNN is trained to relate LIMs to luminosity functions independent of model.  

We want to train our network with maps simulated from a variety of different CO-halo connections, which we can accomplish by generating training data using different parameter values.  We will refer to models generated with the fiducial \citet{Li2016} parameters `fiducial Li' maps and maps generated with random parameters `random Li' maps.  We take as ``priors" on these parameters 10\% of the priors quoted in \citet{Li2016}.

\begin{figure}
	\centering
	\includegraphics[width=0.45\textwidth]{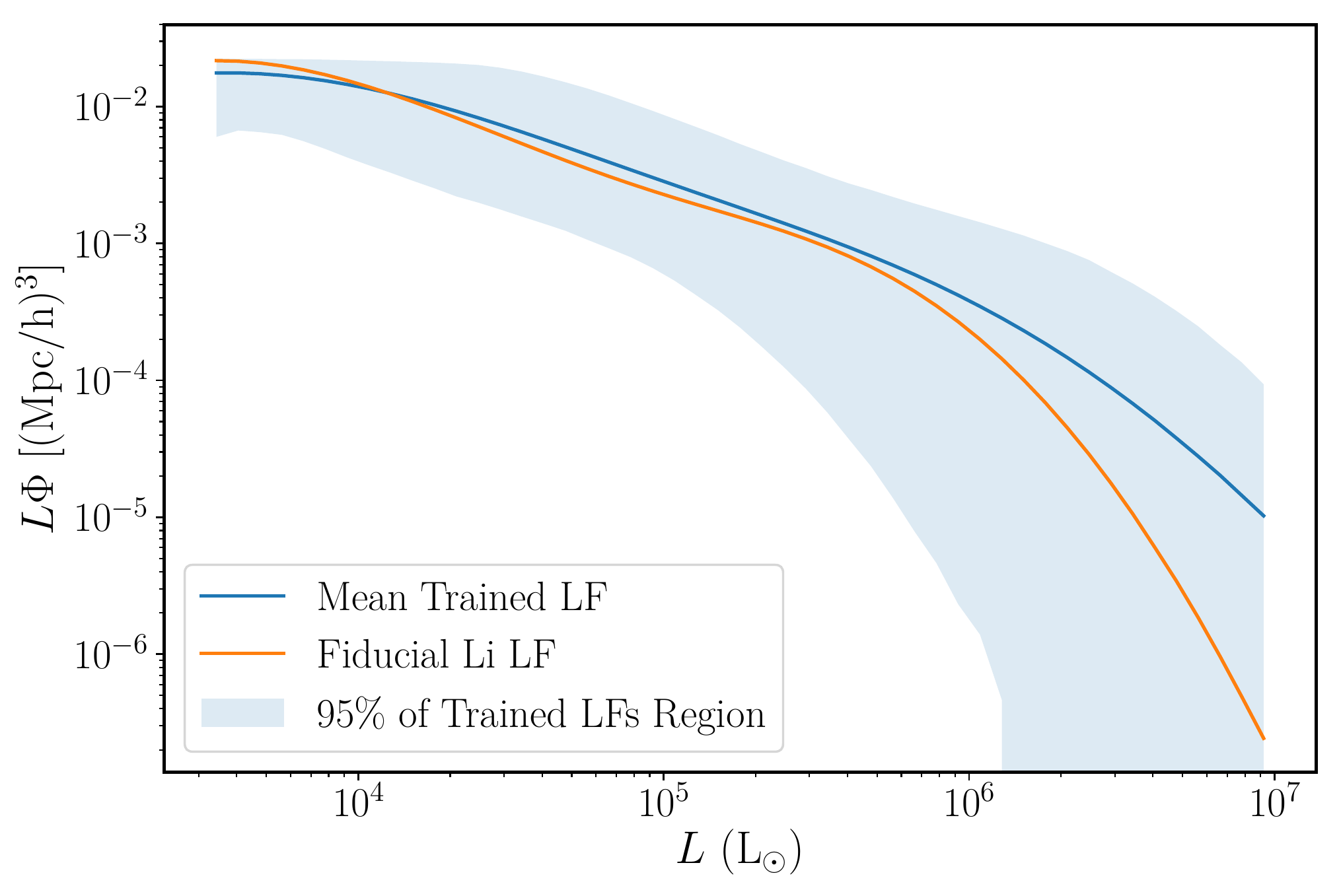}
	\caption{Range of luminosity function values used while training.  The shaded region is the range where 95\% of luminosity function values of a given luminosity fell.  The orange curve shows the 'Fiducial Li' luminosity function and is different than the mean luminosity function used in training.  Note that the 'Fiducial Li' curve is within 95\% region of trained luminosity functions.  It's high value compared to the trained luminosity function range is due to taking the average of logged values and the log-scale of the y-axis.}
	\label{fig:training_lum_funcs}
\end{figure}

We used the publicly available \texttt{limlam\_mocker}\footnote{https://github.com/georgestein/limlam\_mocker} package for line intensity mocks to create the COMAP intensity mocks and corresponding luminosity functions from the 161 halo catalogues, resulting in 5796 possible independent 1.5\textdegree$\times$1.5\textdegree\ COMAP mocks for each choice of parameters/noise/foregrounds.  Figure \ref{fig:training_lum_funcs} shows range of luminosity functions generated for training purposes.  As the luminosity increases, the variance of the luminosity increases as well.

\subsection{Noise and Foregrounds} \label{sec:noise}

\begin{table*}
    \centering
    \caption{Summary of signal, noise, and foreground models used for training and testing.}
    \begin{threeparttable}
    \begin{tabular}{cccc}
      \hline		
      & Model Type & Summary & Used for Training\\
      \hline
      \multirow {4}{*}{CO Signal} & Fiducial Li & \citet{Li2016} CO-halo mass model with fidicual parameter values & No \\
      & Random Li & \citet{Li2016} model with randomly chosen parameter values & Yes \\
      & Padmanabhan & \citet{Padmanabhan2018} CO-halo mass model with fiducial parameter values & No \\
      & Less Bright Sources & Random Li maps chosen by hand to contain $<500$ sources above $L=10^6\ L_{\odot}$ & Yes\tnote{1}\\
      \hline
      \multirow{4}{*}{Instrument Noise} & No noise & No added noise. Hypothetical sample-variance dominated measurement & Yes \\
      & Fiducial Noise & Thermal white noise with fiducial COMAP amplitude & No\\
      & Random Noise & Thermal white noise with amplitude drawn uniformly from \([0, 25.25]\) & Yes \\
      & Geometric Noise & Added white noise around survey edge with amplitude following Eq. (\ref{eq:geoNoise}) & No \\
      \hline
      \multirow{3}{*}{Foregrounds} & No Foregrounds & No added foregrounds, simulates perfect foreground cleaning & Yes\\
      & Fiducial Foregrounds & Point-source emitters drawn following \citet{Muchovej:2009rs} data & No \\
      & Random Foregrounds & \citet{Muchovej:2009rs} with random parameters & Yes \\
      \hline
    \end{tabular}
    \begin{tablenotes}
    \item[1] `Less Bright Sources' maps are a subset of the `Random Li' mocks, so they appear naturally in the training data.
    \end{tablenotes}
    \label{tab:definitions}
    \end{threeparttable}
\end{table*}

To explore how our network might perform in a true analysis we must include instrumental noise and foregrounds in our simulations. In this study we use relatively simple models for both of these effects, considering only thermal instrumental noise and point-source extragalctic foregrounds.

Thermal noise can be modeled by adding an independent Gaussian random number with zero mean and a variance of \(\sigma_{wn}^2\) to each voxel of a LIM.  In the case of COMAP Phase 1 the noise is expected to be \(\sigma_{wn} \simeq 11 \, \mu \rm{K}\) \citep{Li2016} in a $4'$ voxel for a map with 512 frequency channels. To scale this noise to different voxel sizes we make use of 
\begin{equation}
    \sigma_{wn} \propto \frac{1}{\delta \theta \sqrt{\delta v}}.
\end{equation}
We find that, in order to match the COMAP noise properties, our voxels need to have \(\sigma_{wn} \approx 13.88 \, \mu \rm{K}\). As the noise properties of a given survey may not be precisely known \emph{a priori}, we will split our mocks between those with `fiducial noise', i.e. those with the above COMAP noise model, and those with `random noise', in which we assume a different noise amplitude $\sigma_{wn}$.  In either case, a given mock will have a random realization of the given noise model.

A uniform white noise is the only noise source in the ideal case, but in practice many observations include additional, sometimes unknown, sources. One example comes from the scan strategy of the telescope which in general results in different integration times on different pixels, with the central region of the survey generally having a longer integration and therefore a lower noise level. Later when testing our network, we will also include a model of this type of `geometric' noise, where pixels within 5\% of the edge of the survey area have an additional white noise contribution given by \begin{equation}
    \sigma_{\rm{geo}} = \sigma_{\rm{geo,max}} \frac{\rm{max}(d_{\rm{max}} - d, 0)}{d_{\rm{max}}}
    .
    \label{eq:geoNoise}
\end{equation}
In this simple model, \(\sigma_{\rm{geo,max}}\) is the maximum amount of geometric noise we should add at the absolute edge, \(d\) is the shortest distance from the edge of the LIM to a given pixel as a fraction of the the length of the LIM.  We cut off the added noise at a distance \(d_{\rm{max}} = 0.05\).  For our tests we somewhat arbitrarily \(\sigma_{\rm{geo,max}} = 100 \, \mu \rm{K}\).  We leave detailed modeling of specific scan strategies to future work.

Our simulated intensity maps also include radio point sources as possible foregrounds, modeled following \citet{Keating2015}.  We assume the differential source count per unit area per flux is described by the power law,
\begin{equation}
    \frac{dN}{dS} = N_{0} \left( \frac{S}{1 \, \rm{mJy}} \right)^{-\gamma} ,
    \label{eq:FGflux}
\end{equation}
where \(N_{0}\) is a normalization parameter per unit area and flux, \(S\) is the source flux, and \(\gamma\) is the power-law index \citep{Muchovej:2009rs}.  The range of the parameters in this foreground model were found to be \(N_0 = 32.1 \pm 3.0 \, \rm{deg}^{-2} \, \rm{mJy}^{-1}\) and \(\gamma = 2.18 \pm 0.12\).

As we assume our foregrounds are continuum emitters, we assign them pixel-by-pixel rather than voxel-by-voxel.  As with our signal and noise models, we can choose to draw either `fiducial foregrounds' where we assume best fit parameter values from the above foreground models, or `random foregrounds', where we randomly assign parameter values before generating a realization.  In each pixel, we Poisson draw the number of sources based on the expected sources per square degree (thus neglecting large-scale structure correlations in our foregrounds).  We then randomly assign each source an overall with probability set by Eq. (\ref{eq:FGflux}) and a spectral index drawn from the distributions plotted in Figure 3 of \citet{Muchovej:2009rs}.  We can then use this slope and normalization to compute the contribution of the source to each of our frequency

This is a somewhat simplistic and optimistic model of foreground contamination, as it does not include Galactic emission and ignores emission from point sources below the detection threshold of \citet{Muchovej:2009rs}.  However it does capture the essential features necessary for our purposes, in that it results in a map of bright, continuum emission which does not correlate with the large-scale structure of our CO signal. As with the instrument noise, we leave a detailed exploration of foreground emission to future work.

Both the white-noise and foreground additions to the LIMs are randomly generated each time an LIM is used for training.  The same is true of geometric noise, but it is only used for testing purposes after training. We ensure these additions are not static objects in order to help prevent overfitting and give the network more unique LIMs to use for training.

In the above we have described several different choices we can make when modeling the signal and noise in our maps.  For the convenience of the reader we summarize these options in Table \ref{tab:definitions}.  Unless specified, a given map is assumed to use `random noise' and `random foregrounds' models.

In summary, to construct a single CO realization, we:
\begin{itemize}
    \item Generate a dark matter halo catalogues~\ref{sec:sims}
    \item Apply a CO-halo mass model, to paint CO luminosities onto halos (Sec. \ref{sec:CO}).
    \item Use the CO luminosities to produce clean CO LIMs as well as record the true underlying luminosity function (Sec. \ref{sec:CO}).
    \item Generate noise and foreground realization and add to map (Sec. \ref{sec:noise}).
    \item Apply Guassian smoothing of \(4'\) beam to map to match COMAP beam size.
\end{itemize}
When training, we use the `Random Li', `Random Noise', and `Random Foreground' models from Table \ref{tab:definitions}.  The first three steps are done before training as it would take too long to generate new LIMs from scratch each time one was needed.  Noise, foregrounds, and beam smoothing are added during training each time a LIM is looked at.

\begin{figure*}
	\centering
	\includegraphics[width=0.85\textwidth]{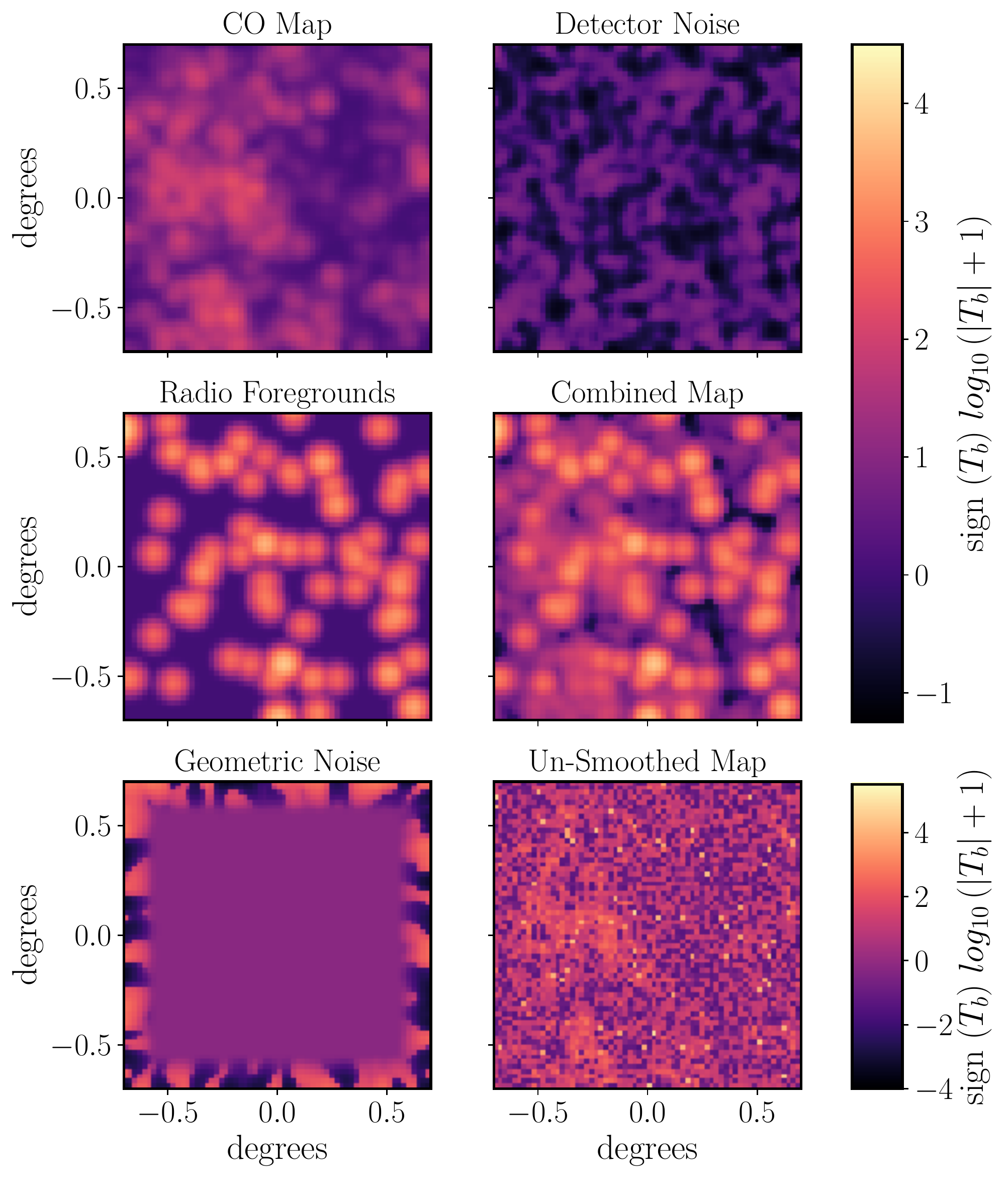}
	\caption{Single slices of different components of our LIMs.  All of the maps have had the log-modulus applied to their intensities (see Eq. \ref{eq:logmod}).  \textit{Top Left}: `Fiducial Li' LIM.  \textit{Top right}: `Fiducial Noise'-only realization.  \textit{Middle left}: Point source foregrounds from `Fiducial Foregrounds' model.  \textit{Middle right}: Sum of previous signal, noise, and foreground maps.  \textit{Bottom left}: `Geometric Noise' realization.  \textit{Bottom right}: Signal+noise+foreground map before beam smoothing.  Note the difference in the color scales between the top two rows and the bottom row.}
	\label{fig:IM_slices}
\end{figure*}

Sample slices of our mock LIMs can be seen in Figure \ref{fig:IM_slices}. We show a `Fiducial Li' LIM as well as maps for white noise, foregrounds, and the sum of all three components.  Both the foregrounds and features of the original random Li LIM are visible in the combined LIM.  For clarity, we also show a realization of the Geometric Noise that we will use for later tests, as well as a version of the combined signal/noise/foreground map before beam smoothing.

\section{Convolutional Neural Network} \label{sec:nn}
The goal of our CNN is to take any LIM as input, and output values of the underlying luminosity function. Due to GPU-memory and training-speed constraints we will downsample our 64x64x100 maps to 64x64x10 by summing groups of 10 voxels along the line of sight.  By lowering the resolution, we can make a larger CNN and train with larger batch sizes.  By adding voxels together, we conserve the total CO luminosity in our LIM, but sacrifice spatial information.  Our forecasts will thus underestimate the constraining power of a network which could handle the full 64x64x512 COMAP data cube.

CNNs are fast NNs designed for classification and regression on images, looking for patterns in a translationally invariant fashion.  Normal images contain two spatial dimensions and a third which stores the intensity of different colors of light (usually three for RGB images).  LIMs behave similarly, albeit with many more spectral channels.  However, a key general advantage of LIMs is that, neglecting noise and foregrounds, the spectral information can be converted directly into a line-of-sight distance.  CNNs typically convolve images in two dimensions, but we can modify this approach to use 3D convolutions and easily make use of the added tomographic information. This one change allows us to make use of the standard framework of CNNs for three-dimensional LIMs. 

After testing a number of CNN architectures we choose a residual learning framework first proposed in \citet{2015arXiv151203385H} for our network architecture. Each layer in these networks (commonly abbreviated as \textit{ResNets}) learns the residual mapping with reference to its inputs instead of directly learning the underlying mapping. This has been shown to improve the training of deep networks with negligible memory or speed tradeoffs. The form of Resnet that we used is a slight modification of the 50-layer network from \citet{2015arXiv151203385H}.  The ordering of the layers was kept the same, but we added an extra residual block in each layer to increase the learning ability of the network without significantly increasing the memory requirement.  Furthermore we modified the end of the network to match our required output and used the maximum number of filters that our GPU would allow.  As mentioned previously, we have also modified the architecture to use three-dimensional convolutions.

This architecture was not designed for this specific problem so we do not believe that it is truly the most optimal possible CNN.  A detailed architecture optimization is beyond the scope of this work, but as ResNets are very common in recent cosmological applications of CNNs we expect the relative results shown here to be representative.  For example, we would expect the decrease in accuracy of prediction when considering noise/foregrounds beyond the fiducial case to hold for more general network designs.

\subsection{Network Architecture} \label{sec:arc}

\begin{figure*}
	\centering
	\includegraphics[width=0.70\textwidth]{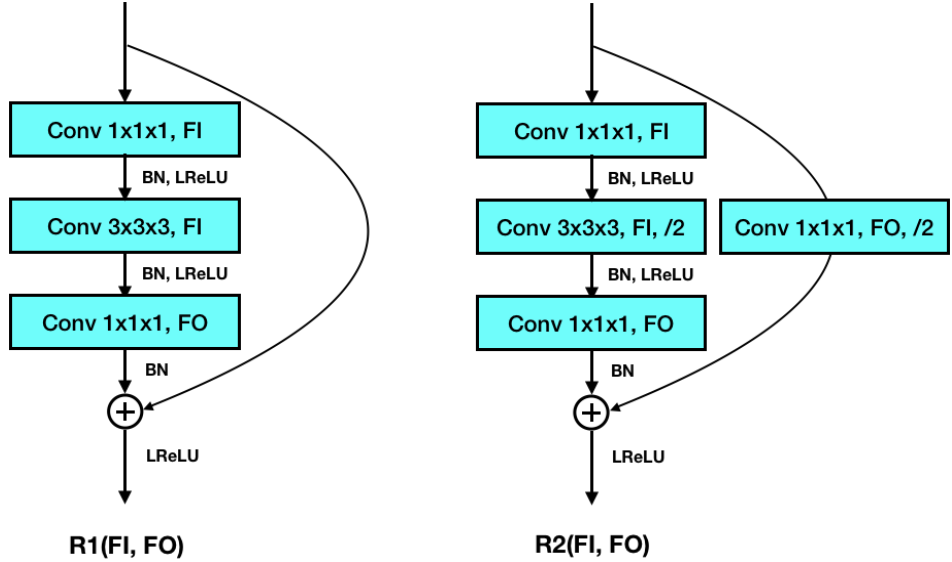}
	\caption{Basic residual block structures used. The teal blocks are used to denote convolutional layers. Both blocks use a bottleneck design of a 1x1x1 convolution followed by a 3x3x3 convolution and ending with a 1x1x1 convolution of varying filter size.  They also include a shortcut which directly adds the input data to the output of the three convolutions in the bottleneck.  See text in Section \ref{sec:arc} for more details.}
	\label{fig:Res_blocks}
\end{figure*}

Here we summarize the modified \citet{2015arXiv151203385H} network we use for this work.  A basic tenet of machine learning is that bigger networks allow one to learn more (i.e. learn a more complicated model or learn a less complicated model better).  However, there are exceptions to this rule.  Adding more layers to a network can lead to degradations of the data flowing through the network and the gradients needed for back propagation, decreasing the accuracy of a CNN.  To get around this, one can make use of a Resnet.  At the heart of the Resnet is the residual block.  A residual block consists of multiple convolutional layers and a shortcut that connects the input directly to the output.  Adding a shortcut helps prevent degradation of network performance because it allows the residual block to function as a small change on top of an identity mapping between input and output.  This means no residual block should give worse results then a previous layer of the network, an issue which large networks without residual blocks face.

After most convolutions, we make use of batch normalization (BN) before applying the leaky rectified linear unit (LReLU) activation function \citep{Maas13rectifiernonlinearities}.  BN helps prevent vanishing and exploding gradients by normalizing the output of a convolution for a given batch of data \citep{2015arXiv150203167I}.  Zero padding is used throughout to obtain the required output dimensionality.  Multiple convolutions are grouped together to form residual blocks.  Our residual blocks make use of a bottleneck design.  The bottleneck replaces two 3x3x3 convolutions with three layers of two 1x1x1 and one 3x3x3 convolutions.  The 1x1x1 convolutions in the bottleneck are responsible for reducing and increasing dimensionality.  Bottleneck designs are used to decrease computational time while retaining network performance.  

In our network we use two different types of residual blocks which can be seen in Figure \ref{fig:Res_blocks}.  Both blocks depend on two parameters: filters-in (FI) and filters-out (FO).  FI and FO are the number of filters the convolutions use at at the beginning and the number of filters the output should have, respectively.  The first residual block on the left, R1(FI,FO), takes input of any size and number of channels.  It then uses a 1x1x1 convolution to change the number of filters to FI and is followed by BN and a LReLU.  Next we use a 3x3x3 convolution with FI channels followed by another BN and a LReLU.  The third convolution is a 1x1x1 convolution that changes the number of filters to FO and is followed by BN.  We then make use of the shortcut and take the original input and add it directly to the output of the third convolution.  We employ an `identity shortcut', named so as we do not modify the input data when using it in the shortcut.  A final LReLU is applied before sending the data to the next layer.  The second residual block is similar, but it changes the dimensionality of the data midway through.  The second convolution in this block uses a stride of 2x2x2 (denoted by /2 in the diagram) to lower the dimensionality. As we changed the shape of the data midway through the block, the shortcut is no longer an identity shortcut.  We apply a 1x1x1 convolution with stride 2x2x2 and FO filters to the shortcut to ensure it matches the output of the rest of the block.

\begin{figure}
	\centering
	\includegraphics[width=0.23\textwidth]{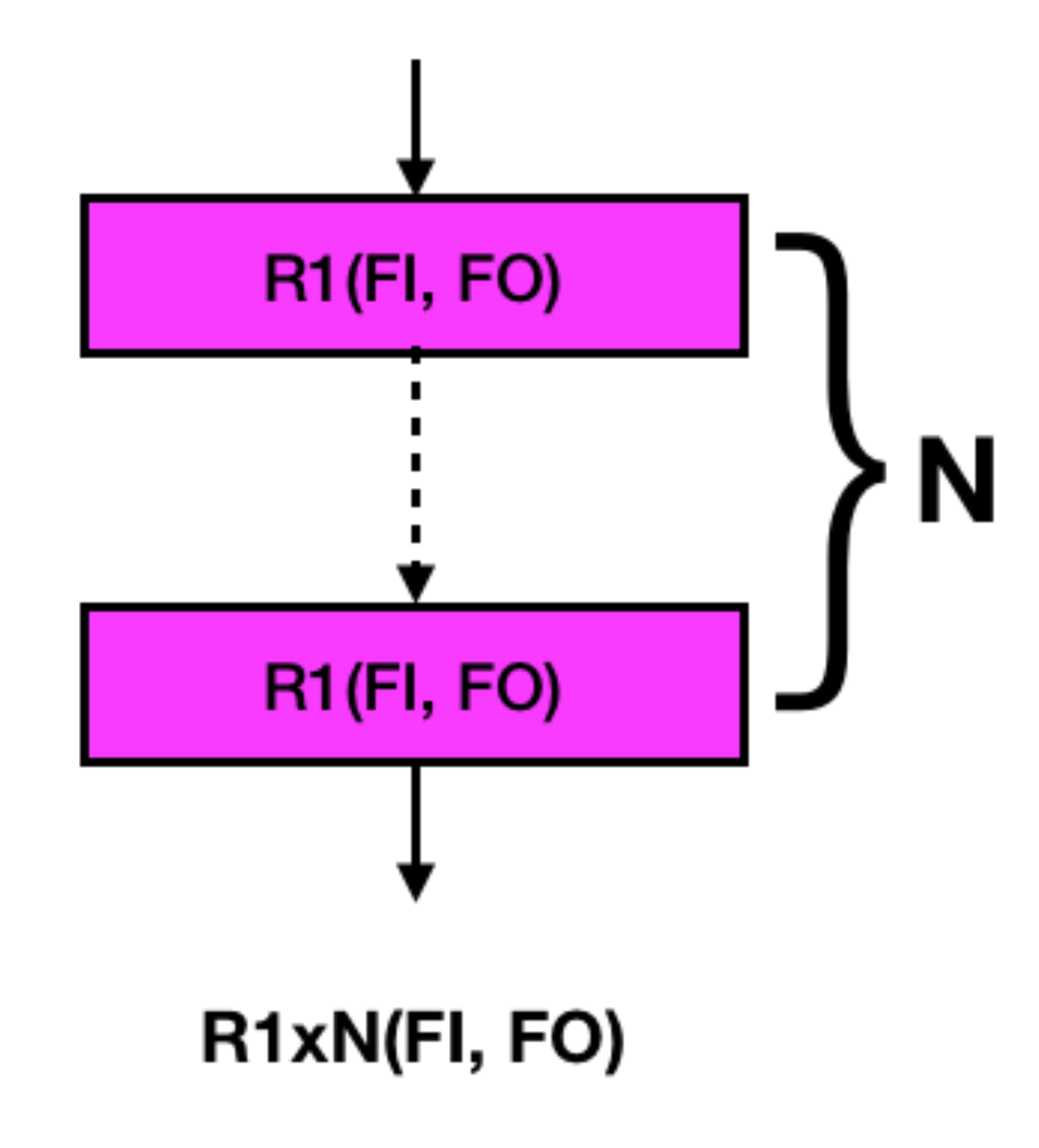}
	\caption{A series of N R1 blocks is defined as an R1xN block.  This structure appears multiple times in our Resnet.  The purple blocks are used to denote R1 blocks.}
	\label{fig:Res_bricks}
\end{figure}

\begin{figure}
	\centering
	\includegraphics[width=0.37\textwidth]{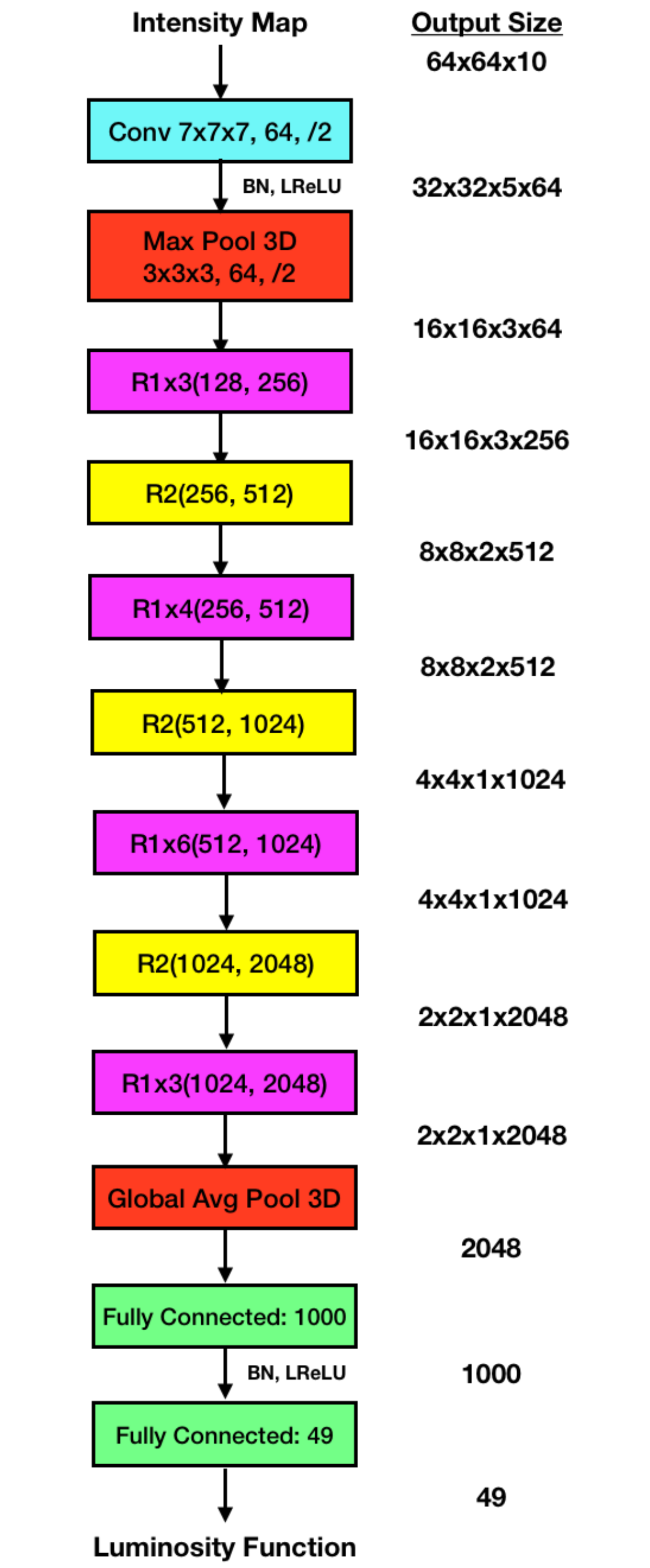}
	\caption{A diagram of our full network.  A 64x64x10 intensity map is converted to 49 different luminosity function values through the use of many convolution layers, two pooling layers and two fully connected layers. Here, the teal blocks represent convolutional layers, the purple represent R1 blocks, the red show pooling layers, the yellow show R2 blocks and green show fully connected layers. }
	\label{fig:resnet}
\end{figure}

We define an R1xN block as N R1 blocks in a row as seen in Figure \ref{fig:Res_bricks}.  With our residual blocks in place, we can now build our full network.  Figure \ref{fig:resnet} displays the full network that accepts a 64x64x10 LIM and outputs 49 values of the luminosity function.  The network starts with a 7x7x7 convolution with 64 filters and a stride of 2x2x2 to reduce the dimensionality by 2.  As usual, we follow the convolution layer with BN and a LReLU.  We then follow up with a 3D max pool with kernel 3x3x3 and stride of 2x2x2.  A max pool layer takes the maximum value within the kernel as output as opposed to a convolution which is effectively the weighted average of the input.  This pool reduces the dimensionality of the data by 2 because of the 2x2x2 stride.  After the pool, we apply three R1 blocks.  We then have three sets of one R2 block followed by multiple R1 blocks, all with the same FI and FO.  The R2 block reduces dimensionality while the R1 blocks increase the depth of the network.  Following the final set of R1 blocks, the data is in the form of a 2x2x1 map with 2048 filters.  On these objects we apply a 3D global average pool which takes the maximum value of the 2x2x1 data for each channel and returns a single value for each channel\footnote{channels for CNNs refer to the number of convolutional filters applied at the last convolutional layer} giving 1-dimensional data.  Second to last, we use a fully connected layer with 1000 neurons which is followed with BN and a LReLU.  Finally, we end with a fully connected layer of 49 neurons with a linear activation function.  Each one of these neurons represents the value of the luminosity function at a specific luminosity.  This architecture can be seen in Table \ref{tab:arch}.

\begin{table}
    \centering
    \caption{Architecture for our Resnet.  Building blocks, as seen in Figures \ref{fig:Res_blocks} and \ref{fig:Res_bricks}, are shown in parentheses with the number of blocks stacked.  Dimensionality reduction is performed by conv2\_1, conv3\_1, conv4\_1 and conv5\_1 by using a stride of 2 for the max pooling or convalution layers.  The total number of training parameters in the network is \(\sim 1.6 \times 10^8\).} 
    \begin{tabular}{c|c|c}
      \hline		
      layer name & output size & layer features \\
      \hline
      conv1 & 32x32x5 & 7x7x7, 64 stride 2 \\ \hline
      \multirow{2}{*}[-10.0 pt]{conv2\_x} & \multirow{2}{*}[-10.0 pt]{16x16x3} & 3x3x3 max pool stride 2 \\ %\cline{3-3}
         & & \( \left( \begin{array}{c}
            1x1x1, 128 \\
            3x3x3, 128 \\
            1x1x1, 256
         \end{array} \right)\) x 3 \\ \hline
      conv3\_x & 8x8x2 &  \( \left( \begin{array}{c}
            1x1x1, 256 \\
            3x3x3, 256 \\
            1x1x1, 512
         \end{array} \right)\) x 5 \\ \hline
      conv4\_x & 4x4x1 &  \( \left( \begin{array}{c}
            1x1x1, 512 \\
            3x3x3, 512 \\
            1x1x1, 1024
         \end{array} \right)\) x 7 \\ \hline
      conv5\_x & 2x2x1 &  \( \left( \begin{array}{c}
            1x1x1, 1024 \\
            3x3x3, 1024 \\
            1x1x1, 2048
         \end{array} \right)\) x 4 \\ \hline
      global\_pool & 2048 & 7x7x7, global avg pool 3D \\ \hline
       fc1 & 1000 & 1000 fully connected \\ \hline
       fc2 & 49 & 49 fully connected \\
      \hline  
    \end{tabular}
    \label{tab:arch}
\end{table}

\subsection{Implementation}

We have made a custom implementation of our Resnet in \texttt{Keras} using the \texttt{TensorFlow} backend \citep{chollet2015keras, tensorflow2015-whitepaper}. We use the default \(\alpha = 0.3\) for our LReLU's. To help with overfitting we apply a dropout rate of \(50\%\) to the second to last layer \citep{2012arXiv1207.0580H, JMLR:v15:srivastava14a}. We obtained our best result when using the Adam optimizer \citep{2014arXiv1412.6980K}.  We used a mean-squared-error loss function for training.

As neural networks work better when data contained within them is similar in magnitude (this includes the input and the data passed between layers), we apply a log modulus function to the values in the intensity map given by
\begin{equation}
    \hat{T} = \rm{sign}(T) \log_{10} \left( |T| + 1 \right) ,
    \label{eq:logmod}
\end{equation}
where \(T\) is the intensity of a given voxel in brightness temperature units.  The sign and absolute value of \(T\) allow us to handle voxels with negative intensity which can come about due to the added white noise.  After transforming the size of the LIM and the intensities within each voxel, we feed the new LIMs into the CNN.  The output of our network was chosen to be 
\begin{equation}
    \mathcal{L}(L) = \log_{10} \left(L \Phi \right) 
\end{equation}
at 49 values of \(L\), where \(\Phi = \frac{dn}{dL}\) is the number density of CO hosting galaxies with luminosity between \(L\) and \(L + dL\).  Returning \(\log_{10}\left(L \Phi\right)\) gives an output that only spans an order of magnitude.  It should be noted that the CNN itself does not know that it is measuring a luminosity function at specific values, only that it is returning 49 ordered numbers.  We consider luminosity bins logarithmically spaced between \(10^{3.5}\) and \(10^{7}\) \(\rm{L}_{\odot}\).

For training and testing our model we start with 5796 `Random Li' LIMs, generated as described in Sec.~\ref{sec:maps}. The LIMs were split into training and validation data with 80\% of the maps being used for training and 20\% used as validation to test our results.  Separating the training and validation data at this point means the network cannot benefit from simply learning the underlying peak-patch halo catalogues.  As noise and foregrounds may be modeled imperfectly in a real observation, we made the conservative choice to use `Random Noise' and `Random Foreground' contamination models. At each step, one of the training signal maps is combined with noise and foregrounds generated with random parameter values.  This allows us to turn our few thousand signal maps into order a million training realizations.  We also make sure to train on noiseless and foregroundless LIMs \(10\%\) of the time to make sure the trained CNN learns to interpret clean LIMs.

\begin{figure}
	\centering
	\includegraphics[width=0.45\textwidth]{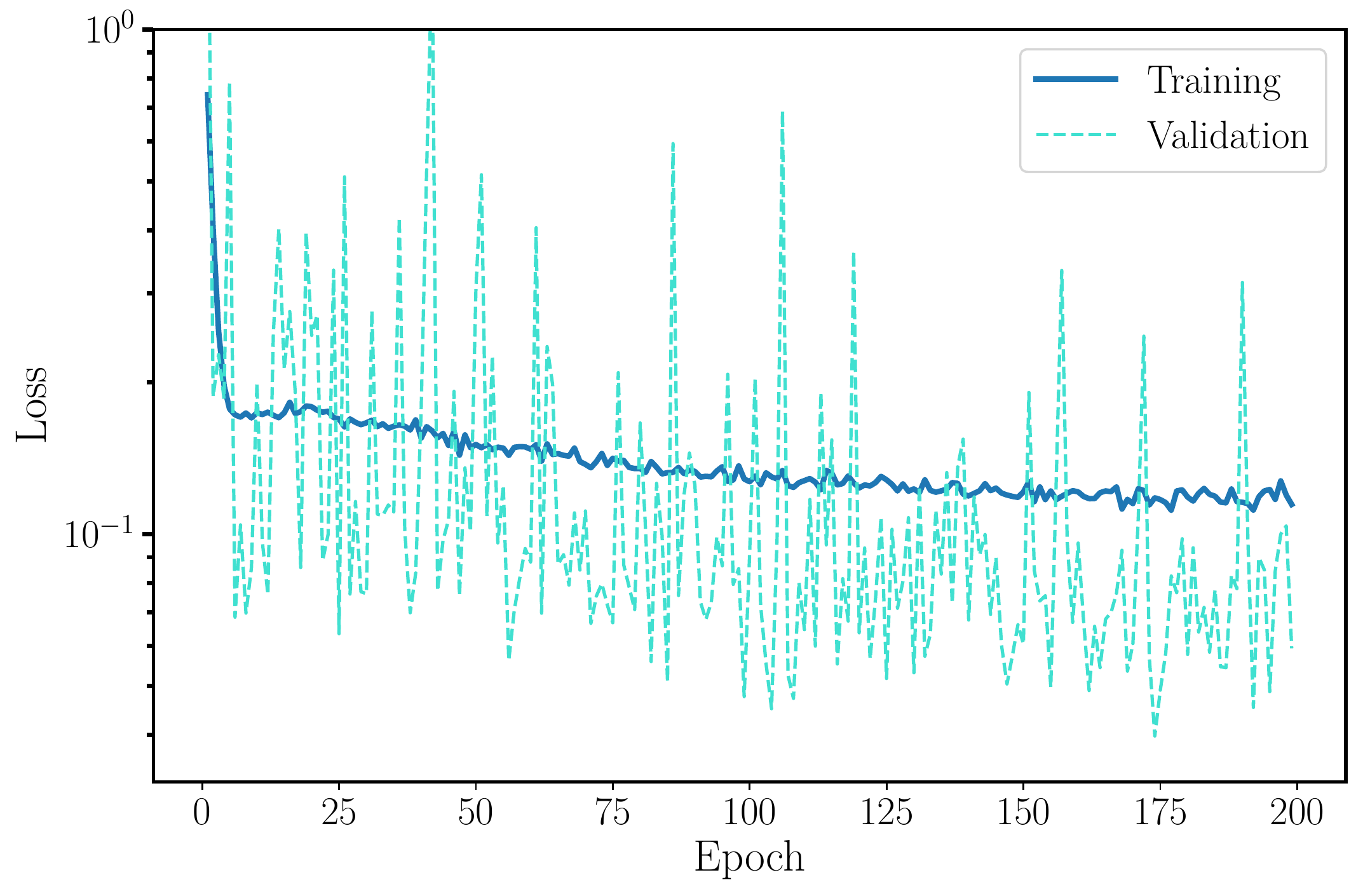}
	\caption{Training and validation loss after each epoch.  The validation loss being less than the training loss is  a result of the dropout in the second to last layer of the network.}
	\label{fig:training_history}
\end{figure}

We trained with batch sizes of 40 LIMs, set by the GPU capacity, and the somewhat arbitrary choice of 150 batches to an epoch for 200 epochs.  We employed 4 Nvidia K80 GPUs with 24 GB each.  The training history of the final CNN is shown in Figure \ref{fig:training_history}. We find that the network does most of its learning within $\sim 2$ epochs.  Learning then slows down dramatically, but the decreasing trend in the loss remains throughout training.  Note that the validation loss can be less than the training loss because of the dropout applied to the second to last fully connected layer of the network.  This is because during training the Resnet only has \(50\%\) of the neurons in the second to last layer working at a time, but during validation or post-training testing \(100\%\) of the neurons are functioning. The Resnet is not as effective without all of its neurons functioning so the loss is often less for validation tests than during training.  

\section{Results} \label{sec:res}

We refer to the trained network as our Resnet.  As a sanity check, we tested it on `Random Li/Noise/Foreground' LIMs which had been rotated by integer multiples of \(90\degree\). Our resnet should be rotationally invariant (at least for \(90\degree\) rotations about an axis parallel to the line of sight of the LIMs), so it should perform similarly on the rotated maps.  After 100 trials, both rotated and unrotated maps had the same average loss and same average variance of the loss within \(<1\%\).

\begin{figure*}
	\centering
	\includegraphics[width=0.90\textwidth]{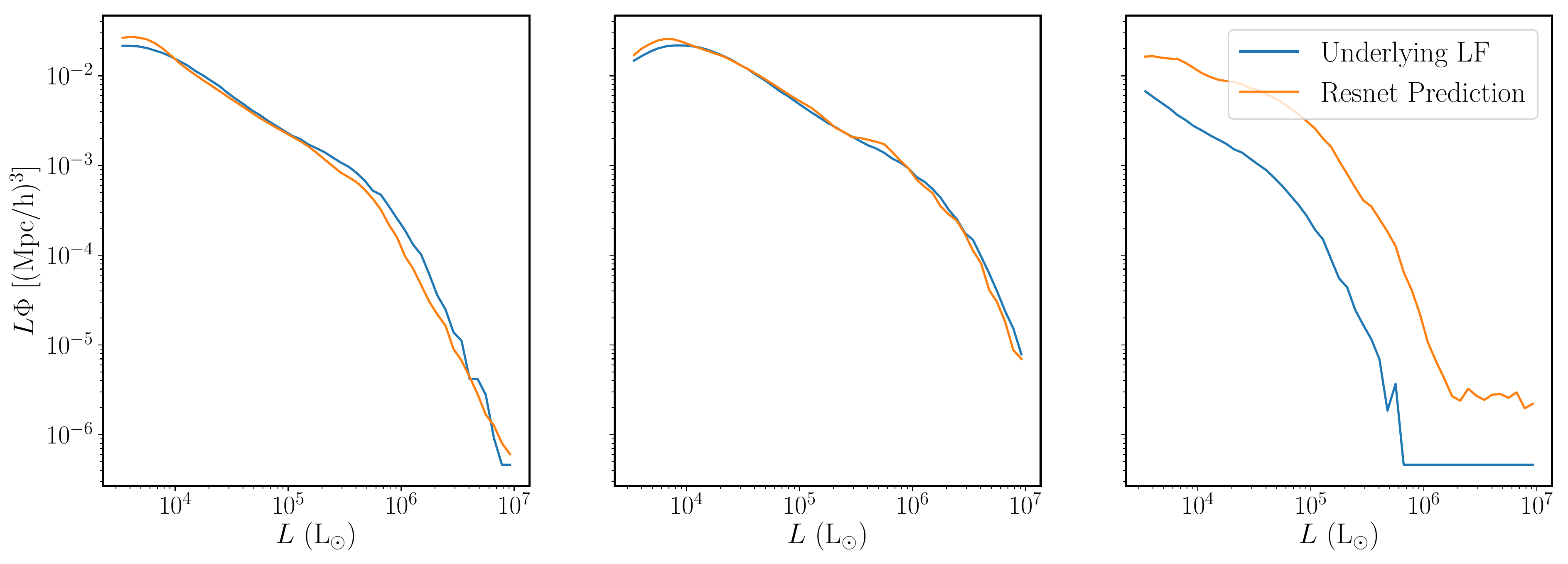}
	\caption{Three luminosity function predictions from the Resnet and the accompanying true underlying luminosity function.  The left figure shows a map with underlying luminosity function that is similar to `Fiducial Li', the middle and right figures show the same for models which differ significantly from the fiducial case, the Resnet performs well in the middle case and poorly in the right case. The three maps were manually chosen from the validation set.}
	\label{fig:guess_range}
\end{figure*}

Example outputs of our Resnet can be seen in Figure \ref{fig:guess_range}.  We show three cases chosen by hand to illustrate different regimes. In the first, the Resnet accurately predicts the luminosity function for a LIM generated from parameters similar to the `Fiducial Li'. The next shows a case where the Resnet performs similarly well for a LIM with a very different luminosity function.  Finally, we show a case where the Resnet fails to accurately reproduce the true luminosity function.  By inspection, we find that the Resnet tends to perform worst when the underlying luminosity function has a low number of bright sources compared to the `Fiducial Li' model.

When training, we know the true luminosity function of a simulated map, so we can use a loss function to assess the network's performance.  However, in a real analysis we would hope to test our trained network on a single data set where the true values are unknown.  In order to estimate how much we could trust the network in this situation, we examine the fractional difference
\begin{equation}
    \delta L \Phi = \frac{L \Phi_{\rm{prediction}}(L) - L \Phi_{\rm{true}}(L)}{L \Phi_{\rm{true}}(L)}.
\end{equation}
as a function of luminosity between the predicted and true quantities.  With a large ensemble of test realizations, we can generate a confidence interval around the true value, which approximates the error bar we would place on a true measurement.  For our figure of merit, we will quote the 95\% confidence interval on $\delta L \Phi$.

Now that we have a trained Resnet, we can study how it performs under different conditions.  We focus on three main scenarios:

\begin{enumerate}
    \item The case the Resnet was designed to handle best: a `Fiducial Li' luminosity function with varying noise and foreground amplitudes.  This models the situation where our fiducial model is close to the truth.
    
    \item A variety of `Random Li' mocks.  Though the Resnet was trained on models in this space, the nature of our priors means that less time is spent on training models that differ significantly from fiducial.
    
    \item Models and contaminants outside of the space of training data.  This accounts for the possibility that the signal on the sky contains aspects not accounted for in the synthetic data.

\end{enumerate}
For each test, we examined a number of maps equal to the number of validation maps used (\(5796 \times 0.2 \approx 1159 \) maps).  It should be reiterated that all of the dark matter catalogs used in this testing step were taken from the set left out of the training data.  Refer to Table \ref{tab:definitions} for a reminder of what effects are included in each test set.

\subsection{Tests on trained data} \label{sec:bt}

In order to assess the performance of our Resnet we can compare to forecasts using analytic methods.  For the simulated COMAP data we are using here, we can compare our forecased constraints on the luminosity function to those from \citet{Ihle2019}.  Their work, which forecast constraints on the underlying luminosity function of a LIM using a joint power spectrum and voxel intensity distribution analysis (PS/VID analysis hereafter), used what we are calling `Fiducial Li' and `Fiducial Noise' models with no foregrounds.

\begin{figure}
	\centering
	\includegraphics[width=0.45\textwidth]{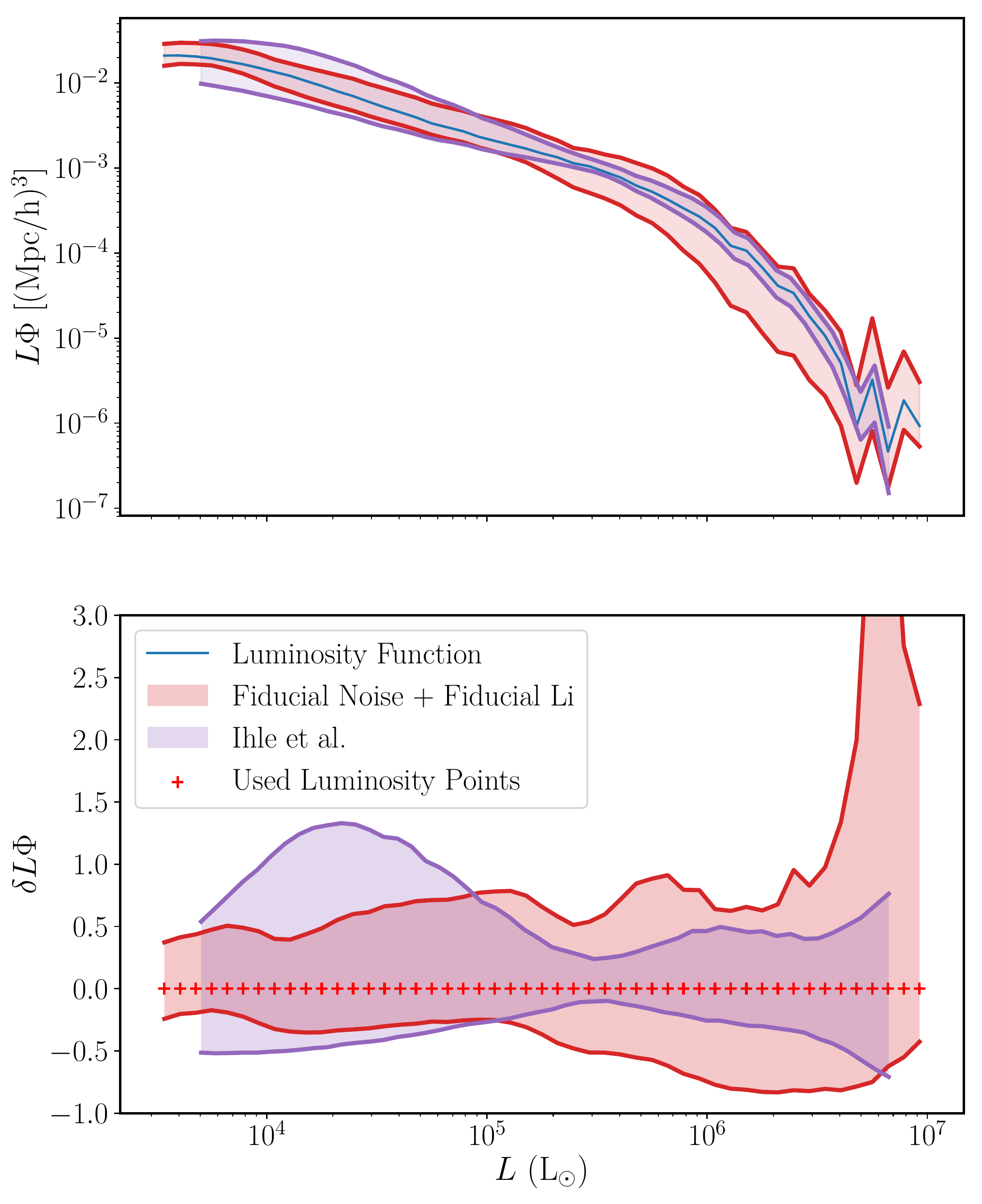}
	\caption{Comparison of 95\% confidence intervals between our Resnet run on `Fiducial Li'/`Fiducial Noise'/`No Foregrounds' test data and the results of \protect\citet{Ihle2019}.  \textit{Top}: \(95\%\) confidence intervals placed directly on top of a `Fidicial Li' lumionsity function.  \textit{Bottom}: Same \(95\%\) confidence intervals on the relative error of the two forecasts.  Red crosses show the luminosities used to train Resnet.}
	\label{fig:Ihle_comparison_full}
\end{figure}

The comparison is shown in Figure \ref{fig:Ihle_comparison_full}.  The contour for the Resnet shows the \(95\%\) confidence interval about the median of the relative errors over the entire set of LIMs tested while the \citet{Ihle2019} \(95\%\) confidence interval comes from their MCMC analysis.  The red crosses show the luminosity bins we used for our Resnet.  Though we are comparing two forecasts using the same models, the forecasts are not exactly equivalent.  The PS/VID forecasts found errors on the \emph{parameters} of the \citet{Li2016} model, then propagate those errors to the luminosity function.  The PS/VID analysis was also able to use the full frequency spectrum of the COMAP data, while our preliminary tests here had to sacrifice much of this information for memory reasons.

With these caveats in mind, the CNN and analytic forecasts appear to perform comparably well, with the Resnet confidence interval actually being smaller at low luminosities.  The Resnet does worst at the highest luminosities, where any given box is expected to have very few sources.  We leave a full comparison between these methods, where the PS/VID MCMC is run on our type of non-parametric model and we have enough computing resources to train on the full COMAP cube, to future work.  Even this rough comparison though is enough to suggest that, in the best case scenario, a CNN can perform similarly to or better than analytic analyses.

As this is the test case that most resembles past work, we will compare all of our upcoming tests to the confidence interval obtained here.

\begin{figure*}
	\centering
	\includegraphics[width=0.85\textwidth]{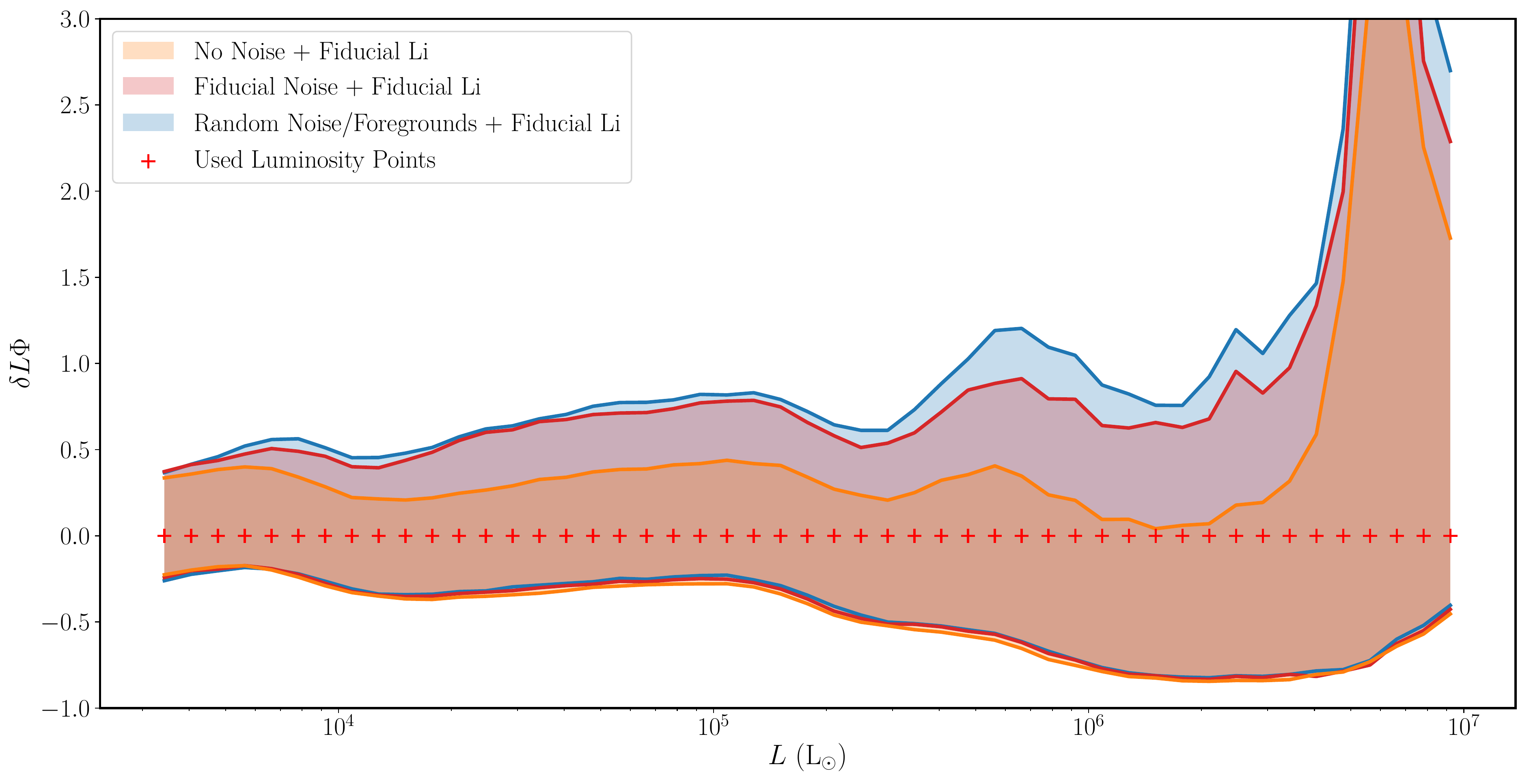}
	\caption{Relative 95\% confidence intervals for test data with no noise/foregrounds (orange), fiducial noise with no foregrounds (red), and random noise/foregrounds (blue).  All LIMs used the `Fiducial Li' signal model.  Scenario descriptions are in Table \ref{tab:definitions}.  The red interval shows the same case from Figure \ref{fig:Ihle_comparison_full}.}
	\label{fig:basic_Li_comparison}
\end{figure*}

Next we examine how our forecasts vary with different noise and foreground levels. Figure \ref{fig:basic_Li_comparison} shows the accuracy of the trained Resnet on `Fiducial Li' LIMs with varying amounts of contamination.  As the noise level of the maps is increased, the quality of the prediction decreases as expected.  However, the Resnet retains significant predictive power even with random-amplitude noise and foregrounds added.  All of the forecasts still begin to fail drastically at the highest luminosities where we start to run out of bright emitters.

\begin{figure*}
	\centering
	\includegraphics[width=0.85\textwidth]{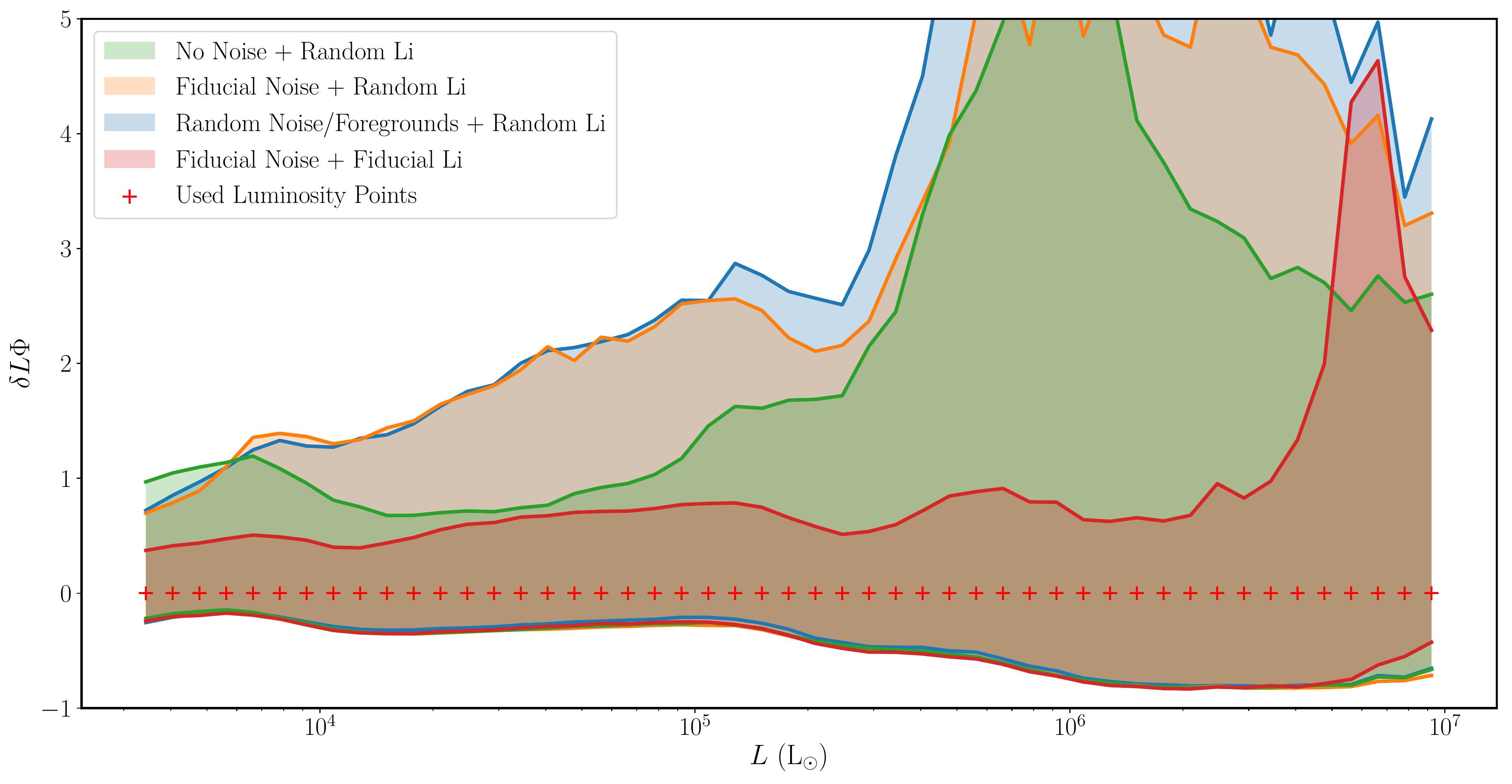}
	\caption{Same as Figure \ref{fig:basic_Li_comparison}, but instead of testing on `Fiducial Li' LIMs it is tested on `Random Li' models.  The red contour from the previous two figures is shown here as a comparison. Red crosses show the luminosities that we trained the Resnet on.}
	\label{fig:small_Li_comparison}
\end{figure*}

Now that we know our Resnet can handle contamination, we want to see how it performs with different signal models. Figure \ref{fig:small_Li_comparison} shows the \(95\%\) confidence intervals for the Resnet results when tested on `Random Li' test LIMs, with same noise scenarios as Figure \ref{fig:basic_Li_comparison}.  As we spend less time training on the different model parameters, these tests are less constraining  than the ones in Figure \ref{fig:basic_Li_comparison}.  In the `Random Li' test set, we expect to periodically encounter cases such as the right panel of Figure \ref{fig:guess_range}, where the Resnet perfoms poorly.  These outliers dramatically reduce the performance of the Resnet, particularly at high luminosity.  This suggests that, while the CNN we have trained here can handle our best-fit model, it would be risky to use in situations where the true data might deviate significantly from expectations.

\subsection{Tests on untrained data} \label{sec:ut}

\begin{figure*}
	\centering
	\includegraphics[width=0.85\textwidth]{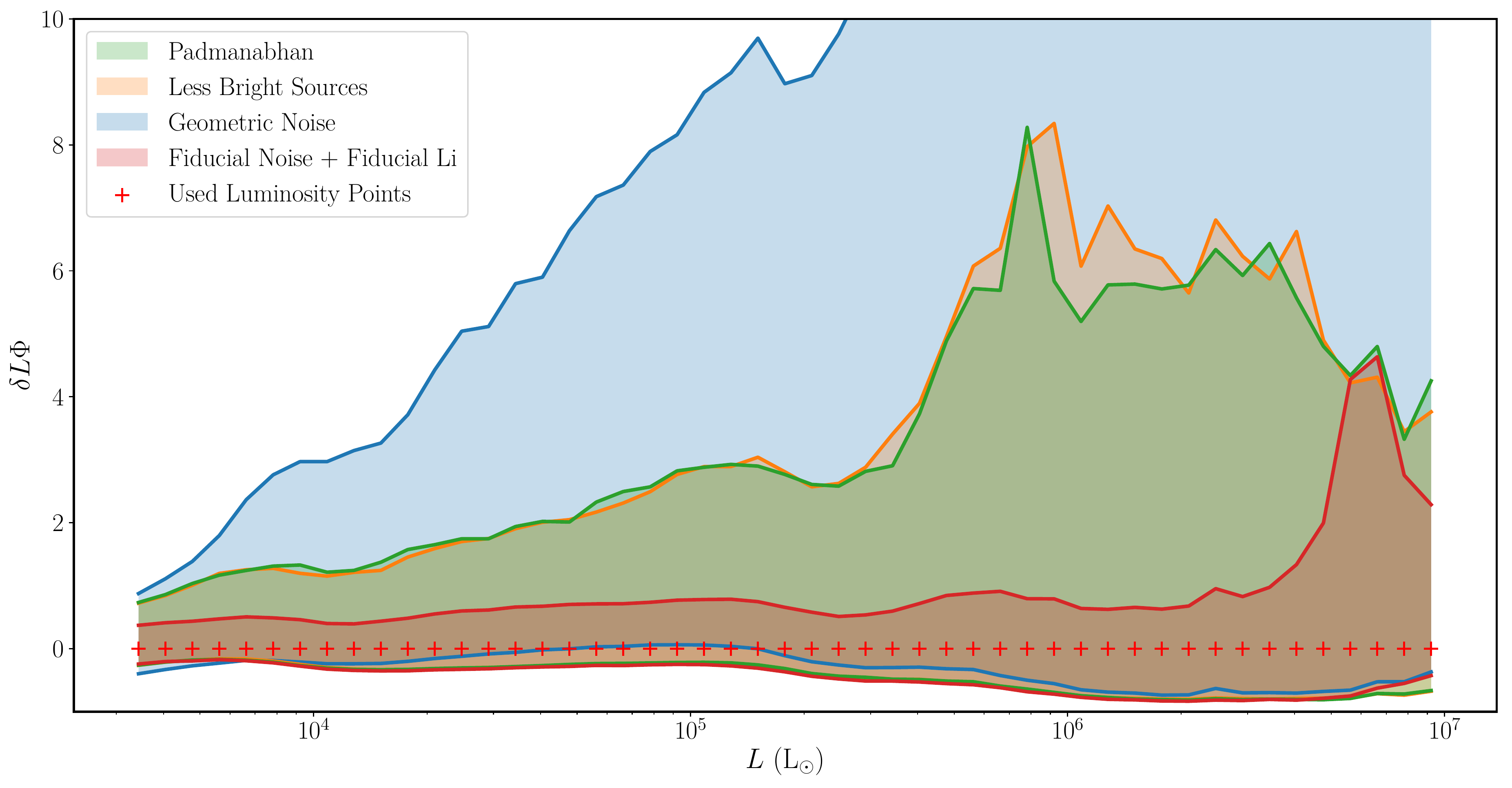}
	\caption{Confidence intervals for test data with effects not included in the training data, including the `Padmanabhan' signal model (green), the hand-chosen `Less Bright Sources' mocks (orange), and the extra `Geometric Noise' contamination (blue). As before, the red contour from previous plots is shown here as a comparison.}
	\label{fig:random_comparison}
\end{figure*}

So far the Resnet was tested on LIMs made from models within the space of the training data.  However, any real observation has a chance of including unexpected effects which are not modeled in advance.  It is generally believed that supervised machine learning, as used in this work, lends itself best to interpolation rather than extrapolation.  Therefore, to see how robust the the trained Resnet is, we must test it on scenarios not included in the training data.  Each of the tests below use `Random Noise' and `Random Foreground' models.  

First, we consider maps with an altogether different model for the underlying luminosity function, specifically that of \citet{Padmanabhan:2017ate}.  The `Padmanabhan' model uses a double power law to relate halo masses to \(L_{\rm{CO}}\), as opposed to the `Fiducial Li' which uses power-law scalings on top of the \citet{behrooziB, behrooziA} star formation rates.  Although both models generally produce similar expected luminosity functions, the detailed shapes are somewhat different with the Padmanabhan model expecting more bright sources.  We tested on only fiducial parameters of the `Padmanabhan' model.  

Next we consider an extra noise source, specifically the `Geometric Noise' model described in Section \ref{sec:noise}.  This is added in addition to the existing `Random Noise' model, and is meant to represent the extra noise around the edges of a survey due to decreased observing time.

Finally, we noted previously that the worst outliers in our `Random Li' sample came from maps with very few bright sources.  To examine this behavior, we consider a hand-curated sample of `Random Li' LIMs that contain fewer than 500 sources above \(L = 10^{6} \rm{L}_{\odot}\).  The \citet{Li2016} best fit parameters lead to more than 1000 sources in this luminosity range, so these maps deviate sinificantly from the fiducial case.  This test considers the possibility that, while the true signal is within the range of the test data, it by chance specifically comes from a regime where the network performs badly.

The confidence intervals for our Resnet on these new scenarios can be seen in Figure \ref{fig:random_comparison}.  We see that, in all cases, the new effects substantially degrade the predictions.  Interestingly, the predictive power is roughly the same for both the `Padmanabhan' and `Less Bright Sources' LIMs.  Geometric noise was something entirely new to the Resnet, so the constraints are sensibly much worse than any other test we consider.

\begin{figure}
	\centering
	\includegraphics[width=0.45\textwidth]{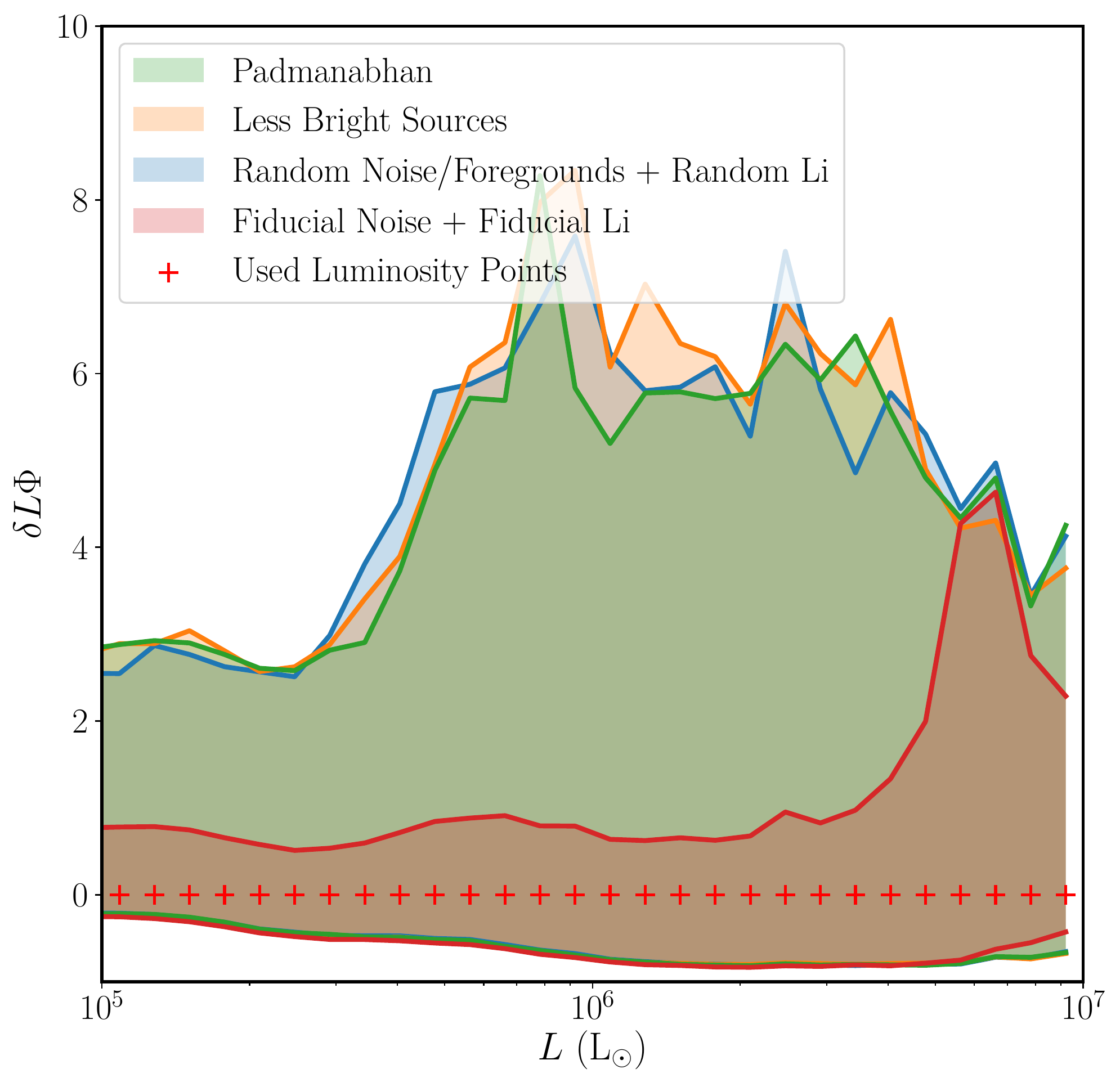}
	\caption{Zoomed in comparison of 95\% confidence intervals for scenarios that fail in a similar fashion.  The `Padmanabhan' (green), `Less Bright Sources' (orange), and `Random Li' cases all yield similar confidence intervals.  As before, the red contour from previous plots is shown here as a comparison.}
	\label{fig:large_comparison}
\end{figure}

Aside from the `Geometric Noise', all of the cases where our Resnet performs poorly seem to yield similar constraints.  Figure \ref{fig:large_comparison} zooms in on the high-luminosity confidence intervals for the `Padmanabhan', `Less Bright Sources', and `Random Li' tests.  The shapes are all extremely similiar.  This implies that the network is failing in a similar fashion when presented with data that is mildly different from the best-case model.

We can use the width of the 95\% confidence interval in different bins to compare between the different cases.   Table \ref{tab:accuracy} compares these values to that of the `Fiducial Li'/`Fiducial Noise' scenario at  \(L = 10^4 \, \rm{L}_{\odot}\).  This allows us to summarize the performance of our Resnet across different scenarios and luminosities.  For comparison, we also include the width of the \citet{Ihle2019}  PS/VID confidence intervals.

\begin{table*}
    \caption{
    Accuracy of the Resnet on different scenarios relative to the accuracy of the `Fiducial Li'/`Fiducial Noise'/`No Foregrounds' case.  Accuracy is measured as the difference between the maximum and minimum values of the \(95\%\) confidence interval at the given luminosities.  All quoted values are relative to the accuracy of the fudicial scenario at \(L = 10^4 \, \rm{L}_{\odot}\).}
    \begin{tabular}{c|c|c|c|c}
      \hline		
       & Scenario & \(10^4 \; \rm{L}_{\odot}\) & \(10^5 \; \rm{L}_{\odot}\) &  \(10^6 \; \rm{L}_{\odot}\) \\\hline
      \multirow{3}{*}{`Fiducial Li' signal} & No Noise/Foregrounds & 0.83 & 0.96 & 1.6\\ 
      & `Fiducial Noise' & 1.0 & 1.3 & 2.2\\ 
      & `Random Noise/Foregrounds' & 1.0 & 1.3 & 2.4\\ \hline
      \multirow{3}{*}{`Random Li' signals} & No Noise/Foregrounds & 1.5 & 1.7 & 7.0\\ 
      & `Fiducial Noise' & 2.0 & 3.2 & 7.9\\ 
      & `Random Noise/Foregrounds' & 1.8 & 3.1 & 9.4\\ \hline
      \multirow{3}{*}{Untrained effects} & `Padmanahban' signal & 1.9 & 3.5 & 7.6\\ 
      & `Less Bright Sources' & 1.7 & 3.4 & 11\\
      & `Geometric Noise' & 3.6 & 8.7 & 35\\ \hline
      \multirow{1}{*}{Other} & Ihle et al. & 2.0 & 1.2 & 0.97\\ \hline
    \end{tabular}
    \label{tab:accuracy}
\end{table*}

\section{Discussion} \label{sec:disc}
From the above results we can now get a sense of how we expect this type of machine learning analysis to perform under different conditions.  The good results for the `Fiducial Li' model indicate that, if all of the contributions to the signal and noise can be well modeled, a CNN can be a useful analysis tool.  However, the other results give reason for caution.  It is not easy to be sure \emph{a priori} whether a map does or does not contain unmodeled effects, so there is not obvious way to tell which confidence interval one should assume around a learned luminosity function.  Therefore, while we have demonstrated the utility of CNNs for this type of measurement, it remains an open question whether this technique can be safely used on real data. 

There are a number of ways our Resnet could be improved for a full analysis.  In this work, we made the simple choice to model the luminosity function as a series of 49 uncorrelated numbers.  However, true luminosity functions tend to vary smoothly, with few sharp features.  Future work could make use of, e.g., a spline model which would retain the non-parametric nature of our forecasts while taking advantage of the smooth nature of the luminosity function.  There also exists a significant amount of space for improving on our network architecture.  The initial downsampling we apply to our maps for memory reasons means that the initial 7x7x7 convolution in our network is actually over is actually over \((1\times1\times10) \times (7\times7\times7) = 3,430\) voxels of the original LIM.  This means that instrument, foreground, and confusion noise can more easily wash out the signal from the underlying emitters.  

We used only a modest amount of computing power for this work, and a future analysis with more resources could relax this downsampling requirement and access more information.  More resources would also enable more training of the network.  We made the somewhat arbitrary decision to stop training after 200 epochs.  However, the loss was still decreasing at this point, suggesting that our constraints would further improve with more training time.  In addition, we did not make a full study of the effects of different architectures.  This work motivates future study of what types of networks are best for studying three-dimensional intensity mapping data.

We noted above that, when handed data that differ slightly but not dramatically from the ideal case, the network seems to fail in a similar manner every time.  Specifically, we obtain the same confidence intervals for the three cases shown in Figure \ref{fig:large_comparison}.  This implies that there is some specific failure happening in our network in a variety of similar situations.  Though it is difficult to tease out the internal logic of a given network, this implies that one may be able to alter our architecture to solve this particular problem.

Interestingly, we see that the Resent nearly always performs better at the lowest luminosities.  This may be due to the underlying range of possible LIMs that we trained it on.  In Figure \ref{fig:training_lum_funcs} we see that the 95\% confidence band starts growing in size above a luminosity of \(10^{5} \rm{L}_{\odot}\) as well.  We leave for future work a study of the effects of these ``prior" choices on the final results.  

We see above that, when allowing for the non-parametric nature of our model and the loss of line-of-sight information, our network performs comparably to the PS/VID analytic forecasts.  This is another area worthy of additional consideration.  Statistics like the power spectrum and VID of a map require human recognition of patterns in data and connection to the underlying physics.  In complicated, highly correlated data sets like we see in LIMs it may be possible for a CNN to recognize additional patterns beyond what we can represent in analytic statistics.  Combined with the fact that our Resnet does best with models near the center of the training range, this perhaps motivates a combined approach.  One can imagine using the confidence intervals from a PS/VID analysis to set the range of training data for a network like ours, then seeing if the trained network can improve on those constraints.  This would have the benefit of allowing the CNN to tease out extra patterns while avoiding some (but not all) of the pitfalls we describe here.

We also need not restrict the methods described here to luminosity function measurements.  LIMs contain information about a wide variety of physics on scales ranging from star forming regions to the large-scale structure.  One could easily imagine training a network like ours to measure cosmological parameters instead of luminosity functions.  There has also been extensive study in the literature on what can be learned from cross-correlations between intensity maps and other tracers \citep[e.g.][]{Lidz2011,Breysse2017,Wolz2017,Fonseca2018,Breysse2019,Chung2019}, which could be studied using our methods by feeding both data sets at once into a CNN.

Another use-case of the results shown here could be for foreground removal.  Though the foreground contamination added to our maps is relatively minor, the network was able to handle it with little loss of accuracy.  Even for maps with more severe contamination, one could apply conventional foreground cleaning to both the real and simulated data and use the network to help account for any residual emission.  This could be particularly useful for cases where the foreground cleaning removes part of the signal in the process \citep{Anderson2018}.  We considered only continuum foreground here, but our same approach may also be useful for separating out interloper emission lines at different redshifts, an effect which is not important for the CO(1-0) maps we consider here but will be very important for maps of several other lines \citep[e.g.][]{Sun2018,Gong2017}

\section{Conclusion} \label{sec:conc}
In this work we have presented the first application of a CNN to determine the underlying luminosity function of line intensity maps.  We considered the example case of CO intensity maps observed by the currently-active COMAP survey.  Under ideal conditions, our Resnet was found to have comparable to better accuracy in predicting the luminosity function as conventional techniques.  This work suggests that, used properly, machine learning could be a valuable tool in extracting astrophysical and cosmological information from intensity mapping data.

However, we also went on to explore some of the weaknesses of these techniques.  We have shown that the accuracy degrades significantly under various conditions that the network was either not trained or insufficiently trained on.  This crucial step has relevance not just to intensity mapping, but to all attempts to use machine learning for cosmological data analysis, and is often missing from past work in the literature.  Though the great potential of neural networks is obvious, this work makes it clear that extreme care must be taken when applying them in this context, as small missing effects can drastically bias the output of neural networks.

Machine learning-based data analysis in cosmology is a field in its infancy.  The proof-of-concept work we present here illustrates both the opportunities and challenges present in the application of these methods.  With proper care, CNNs like our Resnet may play a valuable role in understanding the next generation of experimental data.

\section*{Acknowledgements}
We thank Jared Kaplan, Brice M\'enard, Hamsa Padmanabhan, Kieran Cleary, and Dongwoo Chung for useful discussions.  This research was conducted using computational resources at the Maryland Advanced Research Computing Center (MARCC) and was supported at Johns Hopkins by NASA Grant No. NNX17AK38G.

\bibliographystyle{mnras}
\bibliography{references}

\bsp
\label{lastpage}
\end{document}